\documentclass[11pt,a4paper]{article}
\usepackage[utf8]{inputenc}
\usepackage{amsmath}
\usepackage{authblk}
\usepackage{geometry}
\title{Reply to ``Interpreting Bohm quantum potentials in `Computing quantum waves exactly from classical action' "}
\author{Gábor Vattay}
\affil{Department of Physics of Complex Systems, Eötvös Loránd University, H-1053 Budapest, Egyetem tér 1-3., Hungary\
\texttt{gabor.vattay@ttk.elte.hu}}
\date{\today}

\begin{document}
\maketitle

\begin{abstract}
The recent arXiv posting arXiv:2605.20443 by Lohmiller and Slotine attempts to address the omission of the Bohm quantum potential in their proposed exact equivalence between classical action and the Schrödinger equation. They introduce a position-dependent time transformation to argue that the spatial derivatives of the probability density amplitude vanish for the Feynman kernel. A rigorous mathematical examination of this transformation reveals a violation of the multivariable chain rule. The spatial derivatives do not vanish in the physical reference frame. The mathematical framework presented by the authors remains identical to the well-established semiclassical Van Vleck propagator, which is exact exclusively for quadratic potentials.
\end{abstract}

\section{Introduction}

We thank the authors for their detailed reply and the opportunity to clarify the mathematical foundations of semiclassical wave propagation. In their technical note, the authors acknowledge that for non-quadratic potentials, the divergence of the classical action $\Delta_M \phi_j$ depends on spatial coordinates. To preserve their claim that the Bohm quantum potential $Q_j$ vanishes everywhere, they introduce a local time transformation $t_j'(x,t)$, wherein every fixed position $x$ possesses its own clock. They argue that because the density $\sqrt{\rho_j}$ can be parameterized solely as a function of this new time variable $t_j'$, its spatial Laplacian is strictly zero.

A careful application of multivariable calculus demonstrates that this reasoning is analytically invalid when substituted back into the physical Schrödinger equation.

\section{The Calculus of the Time Transformation}

The Schrödinger equation is formulated in physical space $x$. To determine if a proposed wave function constitutes an exact solution, all differential operators must be evaluated with respect to these physical coordinates.

The authors define a time transformation $t_j'(x,t)$ that explicitly depends on the spatial coordinate $x$ to absorb the local spatial variations of the potential. If the probability density amplitude is expressed as a function of this local clock, $R(t_j'(x,t))$, the physical spatial gradient must be evaluated using the chain rule:
\begin{equation}
\nabla_x R(t_j'(x,t)) = \frac{d R}{d t_j'} \nabla_x t_j'.
\end{equation}
Because the clock speed varies across space to compensate for the non-uniformity of a non-quadratic potential, the spatial gradient of the time coordinate is strictly non-zero ($\nabla_x t_j' \neq 0$). Consequently, the spatial gradient of the density in the physical frame is finite.

Applying the spatial Laplacian yields:
\begin{equation}
\nabla_x^2 R(t_j') = \frac{d^2 R}{d {t_j'}^2} (\nabla_x t_j')^2 + \frac{d R}{d t_j'} \nabla_x^2 t_j'.
\end{equation}
A coordinate transformation cannot physically eliminate the spatial curvature of the probability amplitude in the laboratory frame. When this non-zero Laplacian is substituted back into the Schrödinger equation, the Bohmian quantum potential $- \frac{\hbar^2}{2M} \frac{\nabla_x^2 \sqrt{\rho_j}}{\sqrt{\rho_j}}$ emerges entirely intact. The mathematical requirement for the exactness of the authors' classical construction breaks down.

\section{The Semiclassical Propagator Context}

The authors assert that calculating the wave from purely classical extremal paths yields exact results because they construct a Feynman kernel from a point source, rather than propagating a general wave.

In the formal study of quantum mechanics, constructing a propagator using solely the classical action evaluated along extremal paths, weighted by the classical density, is known as the Van Vleck-Pauli-Morette formula. It is a rigorously proven mathematical theorem that this semiclassical trace formula is exact if and only if the Hamiltonian is at most quadratic in position and momentum. Examples include free particles, uniform gravitational fields, and harmonic oscillators.

In these specific geometric cases, the quantum potential of the kernel genuinely vanishes. The examples provided by the authors, such as the harmonic oscillator, succeed because they fall precisely into this quadratic category where the semiclassical approximation naturally becomes exact.

For non-quadratic systems, such as the Coulomb potential, the semiclassical Van Vleck kernel deviates analytically from the true quantum Feynman kernel. The exact quantum propagator for non-linear potentials inherently requires integrating over the infinite continuum of non-classical paths. This integration is what accounts for the spatial dispersion and generates the quantum potential. By restricting the formulation to classical local extrema, the methodology re-derives the definition of the WKB approximation.

\section{Conclusion}

The proposed local time transformation does not circumvent the need to evaluate spatial derivatives in the physical reference frame. The Bohm quantum potential cannot be mathematically erased by shifting to a spatially dependent clock. The framework detailed in the original paper and the subsequent technical note is indistinguishable from the standard semiclassical approximation and does not constitute a new, exact solution to the Schrödinger equation for arbitrary potentials.

\end{document}